\documentclass[twocolumn,prl,amsmath,amssymb,floatfix,superscriptaddress]{revtex4-2}\usepackage{amsmath}
\usepackage{amssymb}
\usepackage{graphicx}
\usepackage{bm}
\usepackage{color}
\usepackage{hyperref}
\usepackage{graphicx}
\usepackage{upgreek}
\usepackage{scalerel}
\usepackage{multirow}
\usepackage{pgffor}
\usepackage{pdfpages}
\usepackage{mathdots}
\usepackage{float}
\usepackage{silence}
\usepackage{physics} 
\usepackage{lscape}   
\usepackage{color, colortbl}
\usepackage[table]{xcolor}
\usepackage{comment}  
\makeatletter
\AtBeginDocument{\let\LS@rot\@undefined}
\makeatother

\usepackage[normalem]{ulem}

\begin{document}
	\title{Phase-tunable electron transport assisted by odd-frequency Cooper pairs in topological Josephson junctions}
	\author{Jorge Cayao}
	\affiliation{Department of Physics and Astronomy, Uppsala University, Box 516, S-751 20 Uppsala, Sweden}
	\author{Paramita Dutta} 
	\affiliation{Theoretical Physics Division, Physical Research Laboratory, Navrangpura, Ahmedabad-380009, India}
	\author{Pablo Burset}
	\affiliation{Department of Theoretical Condensed Matter Physics, Condensed Matter Physics Center (IFIMAC) and Instituto Nicol\'as Cabrera, Universidad Aut\'onoma de Madrid, 28049 Madrid, Spain}
	\author{Annica M. Black-Schaffer} 
	\affiliation{Department of Physics and Astronomy, Uppsala University, Box 516, S-751 20 Uppsala, Sweden}
	
	\date{\today}
	\begin{abstract}
		We consider  a finite-size topological Josephson junction  formed  at the edge of a two-dimensional topological insulator in proximity to conventional superconductors  and study the impact of Cooper pair symmetries on the electron transport. We find that, due to the finite junction size, electron transport is highly tunable by the superconducting phase difference  $\phi$ across the junction. At zero frequency and $\phi=\pi$, the setup exhibits  vanishing local Andreev reflection and perfect normal transmission due to the interplay of finite junction size and  formation of topological Andreev bound states in the middle of the junction.   We reveal that this striking behavior enables odd-frequency Cooper pairs  to become the only type of pairing  inside the  topological junction that contribute to   transport.  Our paper thus offers a highly tunable detection scheme for odd-frequency Cooper pairs.
			\end{abstract}
	\maketitle
	Engineering superconducting states with unique functionalities has lately triggered great interest, not only due to their unexpected properties but also owing to their potential for quantum technologies \cite{Ac_n_2018}. 
	Odd-frequency superconductivity is an interesting but less explored superconducting state where the basic units, the Cooper pairs, are formed by two electrons at different times. 
	These odd-frequency Cooper pairs are characterized by a wave function, or \textit{pair amplitude}, that  is \textit{odd} in the  relative time, or frequency $\omega$, of the paired electrons \cite{Nagaosa12,Balatsky2017,triola2020role,cayao2019odd}.  
	Initially, the odd-frequency (odd-$\omega$) state was predicted as an intrinsic effect \cite{bere74,PhysRevB.45.13125,PhysRevB.52.1271}, but later it was found that it can be engineered using simple conventional $s$-wave superconductors \cite{PhysRevLett.86.4096,PhysRevB.76.054522}.

Heterostructures based on conventional superconductors have been shown to be the simplest and experimentally most relevant platform for odd-$\omega$ pairing ~\cite{PhysRevLett.86.4096,Kadigrobov01,PhysRevB.76.054522,PhysRevB.86.144506,PhysRevB.87.220506,PhysRevB.92.100507,Lu_2015,PhysRevB.92.205424,jacobsen2016controlling,PhysRevB.94.014504,PhysRevB.95.174516,PhysRevB.95.224502,PhysRevB.96.155426,PhysRevB.97.134523,PhysRevB.98.161408,PhysRevB.101.094505,PhysRevB.101.094506,PhysRevResearch.2.043388,PhysRevLett.125.026802, PhysRevLett.125.117003,PhysRevB.104.054519}. In these systems, breaking the translational invariance at interfaces serves as a source of odd-$\omega$ pairs \cite{cayao2019odd}. In the additional presence of spin mixing fields, e.g., from magnetism or spin-orbit coupling, the even- and odd-$\omega$ pairs can acquire spin-singlet and -triplet symmetries, respectively, even when using conventional superconductors \cite{Eschrig2007,PhysRevB.92.100507,PhysRevB.93.201402,Cayao_2018,PhysRevB.100.104511,PhysRevResearch.2.043193,PhysRevResearch.2.033229,PhysRevB.104.144513,PhysRevB.104.174518,PhysRevResearch.3.L042034}. In all these systems, odd-$\omega$ pairs have provided fundamental understanding of the proximity-induced superconductivity \cite{RevModPhys.77.1321,Nagaosa12,Balatsky2017,cayao2019odd} and have also enabled the entire field of superconducting spintronics \cite{7870d3ff91ed485fa3e55e901ff81c80,EschrigNat15,0034-4885-78-10-104501,yang2021boosting}. Despite all the advances, however, there are still several critical questions that remain unresolved. First, in many cases, the appearance of odd-$\omega$ pairing is also accompanied by an even-$\omega$ component that easily obscures the effect of the odd-$\omega$ part, see e.g., Refs.\,\cite{PhysRevB.76.054522, Nagaosa12,cayao2019odd}. Second, most detection protocols involve observables, such as the local density of states, that do not directly measure pair correlations \cite{PhysRevResearch.2.033229,PhysRevLett.125.117003}. Third, the majority of previous work, including experiments, have mainly used heterostructures with magnetic materials \cite{RevModPhys.77.1321,bernardo15,PhysRevX.5.041021,PhysRevB.101.094505,PhysRevLett.125.117003}, which can easily become  rather challenging as magnetism is detrimental for superconductivity.

With the advent of topological superconductors \cite{flensberg2021engineered}, the spectral bulk-boundary correspondence \cite{PhysRevB.104.165125} predicted a different route for large odd-$\omega$ pairs, thus avoiding some of the previous problems. Topological superconductors have lately attracted great interest not only because they represent another  state of matter but also because they host  Majorana zero modes (MZMs) \cite{flensberg2021engineered}, which are promising candidates for fault tolerant quantum computation \cite{sarma2015majorana}. Perhaps the most appealing way to realize topological superconductivity without magnetism combine  conventional superconductors and two-dimensional topological insulators (2DTI) \cite{PhysRevLett.100.096407,PhysRevB.79.161408}; see also Refs.\,\cite{RevModPhys.82.3045,RevModPhys.83.1057,PSSB:PSSB201248385,Ando13}.  2DTIs have intrinsic strong spin-orbit coupling and host metallic 1D edge states which only experience Andreev reflections in normal-superconductor (NS) 2DTI-based heterostructures \cite{Sullivan_2012,Hart_2014,Bocquillon_2016,Bocquillon17,Deacon_2017,Sajadi_2018,Fatemi_2018}. These Andreev processes have recently been shown to generate odd-$\omega$ spin-triplet pairs \cite{PhysRevB.92.100507,PhysRevB.92.205424,PhysRevB.96.155426,PhysRevB.97.134523}, notably without the presence of magnetic fields, but still accompanied by even-$\omega$ pairs that have so far challenged the identification of odd-$\omega$ pairing. As a consequence, despite large interest in odd-$\omega$ pairs, the generation of pure odd-$\omega$ pairs and their detection still represent two open critical problems.

\begin{figure}[!t]
		\centering
		\includegraphics[width=0.99\columnwidth]{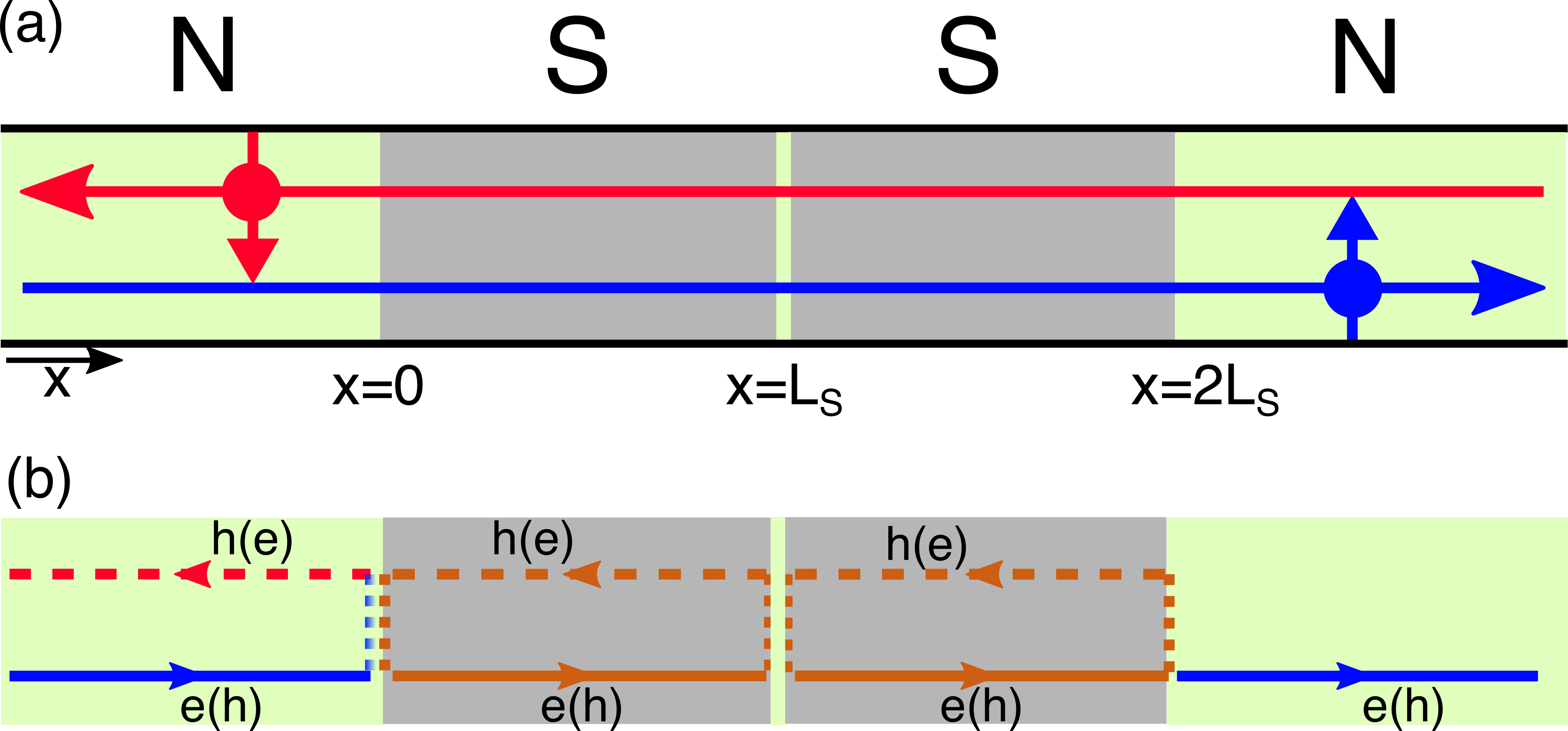}
		\caption{(a) Sketch of N-SNS-N junction  at the 1D metallic edge of a 2DTI, with N (S) regions colored  green (gray) and a middle N region of vanishing length.  The S regions have finite lengths ($L_{\rm S}$) and order parameters, with $\Delta$ and $\Delta\,{\rm e}^{i\phi}$ for the left and right S region, respectively.  Horizontal red and blue arrows depict the counter propagating edge modes  carrying opposite spin (vertical arrows).
		(b) An injected electron (hole) from the leftmost N can be Andreev reflected (dashed lines) or transmitted (solid lines) into the rightmost N as an electron due to the helical nature of the 2DTI. }
		\label{Fig1} 
	\end{figure}

In this paper, we consider a finite Josephson junction formed at the edge of a 2DTI and identify the  Cooper pair symmetries responsible for the electron transport observables.  In particular, we focus on N-SNS-N junctions with a very short middle N region and finite S regions, where  superconductivity is proximity-induced from conventional superconductors and where an external flux controls the superconducting phase difference ($\phi$) between them, see Fig.\,\ref{Fig1}(a).  Interestingly, we obtain vanishing local Andreev reflection and perfect normal transmission at frequency $\omega=0$ and phase $\phi=\pi$ as a result of the interplay between the formation of a pair of topological Andreev bound states (ABSs) and the finite junction size. 
	We reveal that this behavior is accompanied by the emergence of \emph{only} odd-$\omega$ mixed spin-triplet pairs in the middle of the junction, thus unveiling the Cooper pair symmetry that determines transport in this special regime.  Since the helicity constraints make  the Andreev reflection and normal transmission directly set the local and nonlocal conductances, respectively, we conclude that electron transport provides an excellent and controllable way to detect odd-$\omega$ pairs in 2DTI Josephson junctions. 
	
	\emph{2DTI Josephson junction}.---We consider a N-SNS-N junction at the  edge of a   2DTI with N being normal-state regions and S  being  regions in contact with conventional spin-singlet $s$-wave superconductors, see Fig.\,\ref{Fig1}(a).  For simplicity, we assume a middle N region of vanishing length   and S regions of the same length $L_{\rm S}$, with the interfaces located at $x=0, L_{\rm S}, 2L_{\rm S}$. This effectively 1D system is then modeled by the Bogoliubov-de Gennes (BdG) Hamiltonian,
	\begin{equation}
		\label{H2DSC}
		\begin{split}
			H_{\rm BdG}&= v_{\rm F}p_{x}\tau_{z}\sigma_{z}-\mu\tau_{z}+{\bf \Delta}(x)\tau_{x}\,,
		\end{split}
	\end{equation}
	in the basis $ \Psi(x)=
	(\psi_{\uparrow},
	\psi_{\downarrow},
	\psi_{\downarrow}^{\dagger},
	-\psi_{\uparrow}^{\dagger}
	)^{T}$, where $T$ denotes the transpose operation and $\psi^{\dagger}_{\sigma}(x)$ adds an electron with spin $\sigma=\uparrow,\downarrow$ at position $x$ along the edge.  The first term in Eq.\,(\ref{H2DSC}) represents the 1D metallic edge of a 2DTI \cite{PhysRevLett.95.226801,PhysRevB.79.161408,0034-4885-75-7-076501,culcer2020transport}\footnote{In order to obtain perfect 1D metallic edges in 2DTIs, it is beneficial to have large width sample such that the width is much larger than the localization length of the edge states, such as in Ref.\,\cite{vlad15}.}, where the spin quantization direction is along the $z$-axis, $p_{x}=-i\hbar\partial_{x}$, $v_{\rm F}$ is the Fermi velocity of the edge state, and $\sigma_{i}$  ($\tau_{i}$) is the $i$th Pauli matrix in spin (Nambu) space. Without loss of generality we  assume $\hbar=1$, $v_{\rm F}=1$ \cite{culcer2020transport}. Next, $\mu$ is the chemical potential and the last term,  ${\bf \Delta}(x)$, is  the induced superconducting order parameter at the edge of the 2DTI, which is ${\bf \Delta}(x)=0$ in the N regions, $ {\bf \Delta}(x)=\Delta$ in the left S, and $ {\bf \Delta}(x)=\Delta {\rm e}^{i\phi}$ in the right S.  Here, $\phi$ is the superconducting phase and $\Delta$ the order parameter amplitude which introduces the superconducting coherence length $\xi=v_{\rm F}/\Delta$. We note that a finite $\phi$, combined with the 2DTI helicity, enables the formation of topological ABSs and the emergence of MZMs at $\phi=\pi$ \cite{PhysRevB.79.161408,PhysRevB.99.100504}. As we will see, these topological properties are strongly affecting  the behavior of the Josephson junction studied here.	
	
	We are interested in studying the induced Cooper pair symmetries and their impact on transport across the N-SNS-N junction in Fig.\,\ref{Fig1} and modeled by Eq.\,(\ref{H2DSC}). In the following, we first determine the equilibrium transport based on scattering processes, which we then employ to  identify  the induced Cooper pair symmetries and their impact on transport signatures. To highlight the physics, we here discuss the results and present the detailed calculations in the Supplemental Material  \cite{SM}.

	\begin{figure}[!t]
		\centering
		\includegraphics[width=.48\textwidth]{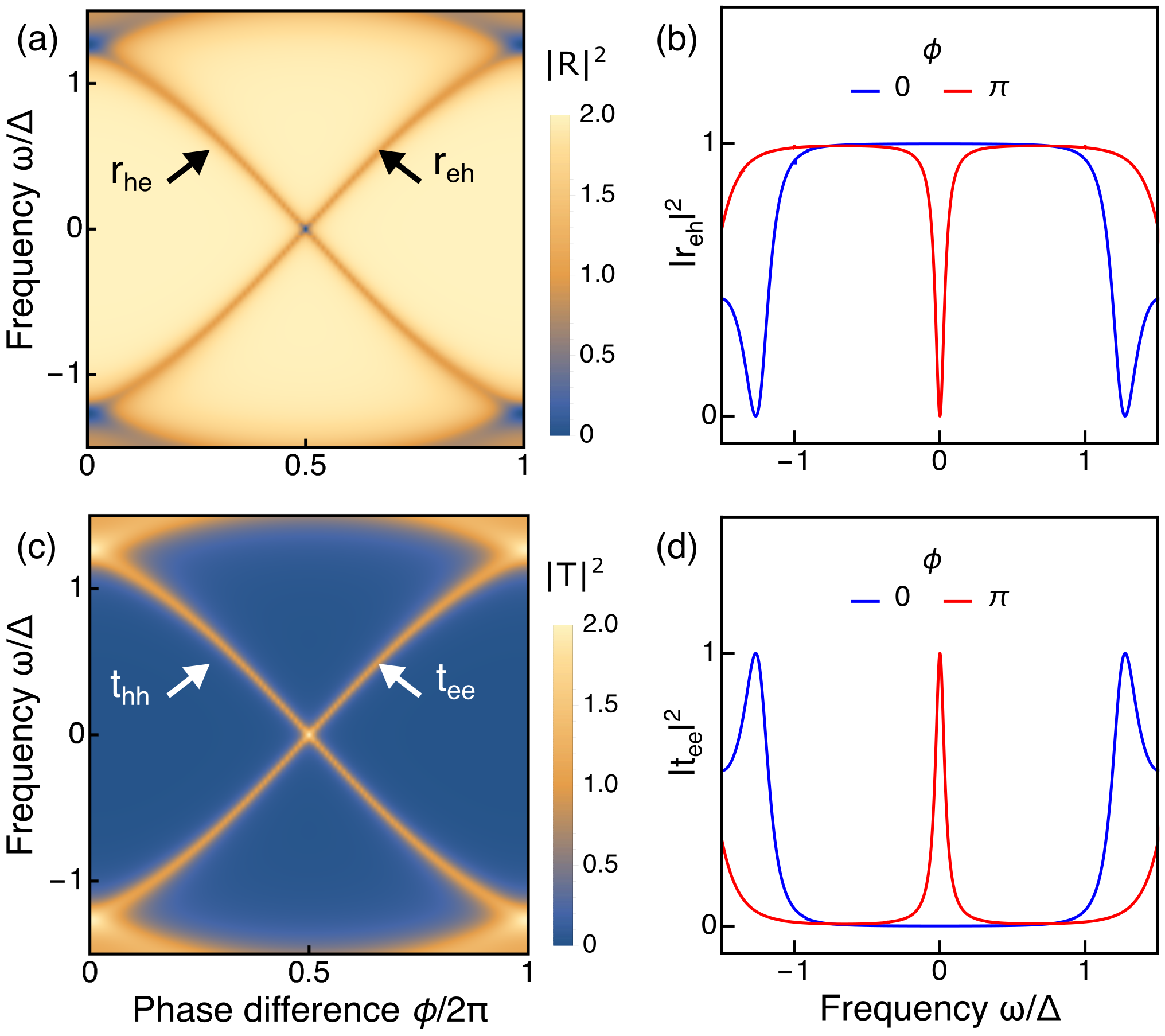} 
		\caption{Scattering probabilities. 
			(a) Total Andreev reflection probability $|R|^{2}=|r_{eh}|^{2}+|r_{he}|^{2}$ due to the incidence of an electron ($r_{eh}$) and a hole ($r_{he}$) from the leftmost N region as a function of $\omega$ and $\phi$; black arrows indicate the phase-dependent branches from $r_{eh (he)}$. 
			(b) Andreev probability, $|r_{eh}|^{2}$, as a function of frequency at $\phi=0,\pi$. 
			(c,d) Same as (a,b) but for the transmission $|T|^{2}=|t_{ee}|^{2}+|t_{hh}|^{2}$ to the rightmost N region. 
			Parameters: $\mu=\Delta$, $L_{\rm S}=2\xi$. }
		\label{Fig2} 
	\end{figure}
	\emph{Transport signatures}.---To begin, we identify the microscopic scattering processes responsible for reflections on one side of the Josephson junction and transmissions across it, which define the measurable local and nonlocal conductances of the junction, see Fig.\,\ref{Fig1}. Due to the 2DTI helicity, normal reflections and crossed Andreev reflections are forbidden \cite{PhysRevB.82.081303,PhysRevB.88.075401,PSSB:PSSB201248385}: a right moving electron (hole) from the leftmost N region  [blue solid arrow in Fig.\,\ref{Fig1}(b)]  can only be   reflected  as a hole  (electron) at the NS interface at $x=0$. This process, known as local Andreev reflection, results in the creation (annihilation) of a Cooper pair in S and we characterize it by the amplitude $r_{eh (he)}$. The right moving electron (hole) can also be  transmitted, ultimately all the way to the  rightmost N, but only as an electron (hole) [blue solid arrow in  Fig.\,\ref{Fig1}(b)]; a process known as normal transmission and here characterized by $t_{ee (hh)}$. 
	As a consequence, only the local Andreev reflection $r_{eh (he)}$ and normal transmission $t_{ee (hh)}$ determine local and nonlocal conductances, respectively, across the junction. 
	In particular, at zero temperature the local conductance in the leftmost N region after applying a bias voltage $V>0$ ($V<0$) is related to the Andreev reflection as $\sigma_{\rm LL}=(e^{2}/h)(1+|r_{eh(he)}|^{2})$ per spin channel. The nonlocal conductance, measured in the rightmost N region for the same bias, is related to normal transmission as $\sigma_{\rm LR}=(e^{2}/h)|t_{ee(hh)}|^{2}$ per spin channel. In the following, it is sufficient to study $r_{eh (he)}$ and $t_{ee (hh)}$, which we obtain by matching the scattering states of the system at all its interfaces  \cite{SM}.

	In Fig.\,\ref{Fig2}(a,c) we plot the total local Andreev reflection  $|R|^{2}=|r_{eh}|^{2}+|r_{he}|^{2}$ and the total transmission $|T|^{2}=|t_{ee}|^{2}+|t_{hh}|^{2}$ as a function of frequency $\omega$ and phase difference $\phi$. In Fig.\,\ref{Fig2}(b,d) we additionally show $|r_{eh}|^{2}$ and $|t_{ee}|^{2}$ as a function of $\omega$ at $\phi=0,\pi$.  The first observation is that these scattering processes acquire a strong phase dependence for  $|\omega|<\Delta$. In fact, $R$ and $T$ develop regions that strongly depend on $\phi$ and follow a cosine profile that reflects the formation of the pair of  topological ABSs  at $x=L_{\rm S}$, known to emerge in 2DTI Josephson junctions \cite{PhysRevB.79.161408}.  The signature of these ABSs in  $R$ and $T$ occurs because the scattering amplitudes are heavily dependent on the poles of the scattering matrix, which gives the conditions for the formation of bound states in a scattering system, see, e.g., Refs.\,\onlinecite{Beenakker:92,PhysRevB.96.155426,PhysRevResearch.2.022019}. The key property of  topological ABSs is that, due to the helicity, they develop a zero-frequency crossing at $\phi=\pi$, which is protected by  the conservation of the total fermion parity and  signals the emergence of MZMs.  We further note that even when the middle N region of the topological Josephson junction in Fig.~\ref{Fig1} has a finite length and hosts additional pairs of ABSs within the gap, the low-energy results presented here  remain unchanged~\cite{PhysRevLett.110.017003, PhysRevLett.112.077002, PhysRevB.89.205115, PhysRevB.96.155426}. This property can be also seen by noting that each ABS emerges from a distinct Andreev process, $r_{eh}$ and  $r_{he}$ [black arrows in Fig.\,\ref{Fig2}(a)], which gives rise to a protected zero-frequency crossing at   $\phi=\pi$ [the same holds for $|T|^{2}$  in (c)].  The fact that local Andreev reflection and normal transmission  capture  the topological ABSs  is a remarkable property, as it implies that they can be easily detected in local and nonlocal conductances.

	The second observation is that the strong phase dependence of the total Andreev reflection $R$ and transmission $T$  reveals a very distinct behavior at $\phi=\pi$ when $\omega=0$. This regime is here of particular interest as it is a key property of  topological ABSs \cite{PhysRevB.79.161408}. To further understand the very distinct behavior of $R$ and $T$ in this regime, we derive the analytical expressions for the Andreev and transmission amplitudes at $\omega=0$,  
	\begin{equation}
		\label{EQRATEE}
		\begin{split}
			r_{eh}(\omega=0)&=\frac{1}{2i}\frac{(1+{\rm e}^{-i\phi}){\rm sinh}(2 L_{\rm S}/\xi)}{{\rm sinh}^{2}(L_{\rm S}/\xi){\rm e}^{-i\phi}+{\rm cosh}^{2}(L_{\rm S}/\xi)}\,,\\
			t_{ee}(\omega=0)&=\frac{1}{{\rm sinh}^{2}(L_{\rm S}/\xi){\rm e}^{-i\phi}+{\rm cosh}^{2}(L_{\rm S}/\xi)}\,,
		\end{split}
	\end{equation}
	with   $r_{he}(\phi)=r_{eh}(-\phi)$ and $t_{hh}(\phi)=t_{ee}(-\phi)$.  At $\phi=0$, the Andreev reflection  is in fact constant for frequencies $|\omega|<\Delta$, becoming $r_{eh}=1$ when $2L_{\rm S}\gg\xi$, while  $t_{ee}$ instead vanishes in this regime, see blue curves in Fig.\,\ref{Fig2}(b,d).  In contrast, at $\phi=\pi$, the Andreev reflection and transmission develop, respectively,  a dip reaching $r_{eh}=0$ and a resonant peak $t_{ee}=1$ at $\omega=0$, surprisingly, for all $L_{\rm S}$  \footnote{For S regions of different lengths,  Eqs.\,(\ref{EQRATEE}) acquire more complicated forms but still develop a dip/peak for Andreev/transmission probabilities, albeit their height and width reduce and increase, respectively, see \cite{SM} for details.}. 
	
	The origin of the unusual behavior of the scattering processes, expressed both in Fig.~\ref{Fig2} and Eq.~\eqref{EQRATEE}, is directly connected to the special structure of the finite topological Josephson junction under study. First, due to the helical nature of the 2DTI edge states,  the junction  hosts a pair of topological ABSs for any $L_{\rm S}$ \cite{PhysRevB.79.161408}. 
	Moreover, owing to the finite length, $2L_{\rm S}$, the system traps discrete levels, which, however, only emerge for $|\omega|>\Delta$ and therefore do not affect transport within the gap. 
	Thus,  the  Andreev reflected hole (electron) in the  leftmost N, characterized by $r_{eh (he)}$, actually carries all the information of  a  full closed cycle within the finite junction, including transmission across $x=L_{\rm S}$ and  Andreev reflection at $x=2L_{\rm S}$. 
	As a consequence, the numerator of $r_{eh (he)}$ acquires a $(1+{\rm e}^{-i\phi})$ phase dependence,   making it vanish at $\phi=\pi$.
	Moreover, we have verified that in the absence of the rightmost N region, i.e., when the right S region is instead semi-infinite, the effect of $\phi$ on $r_{eh}$ becomes a global complex phase, thus leaving no information about the topological ABSs  in the Andreev probability $|r_{eh}|^{2}$.   Hence, the interface at $x=2L_{\rm S}$, which defines the finite length of the S regions, is absolutely necessary to reveal the formation of topological ABSs at $x=L_{\rm S}$ in the scattering probabilities and thus also in the conductance. 
	
	To summarize, local and nonlocal conductance  measurements, set directly by the  local Andreev reflection and  normal transmission, respectively, are enough to detect the formation of  topological ABSs, thus also revealing the presence of MZMs at $\phi=\pi$.  
Since equilibrium transport  in superconducting junctions involves the  transfer of Cooper pairs for subgap frequencies, it is natural to next ask about the symmetry of the Cooper pairs that creates these striking transport features.

	\emph{Induced odd-$\omega$ pairs}.---We next investigate the symmetries of the induced  Cooper pairs in the junction by analyzing the superconducting pair amplitudes. In general, the pair amplitudes are obtained from the  electron-hole (eh) part of the retarded Green's function $G^{r}$  \cite{cayao2019odd,triola2020role}, which satisfies $[\omega-H_{\rm BdG}(x)]G^{r}(x,x',\omega)=\delta(x-x')$, where $H_{\rm BdG}$ is a matrix in spin and Nambu spaces given by Eq.\,(\ref{H2DSC}). To obtain   $G^{r}$ we follow the method  based on  scattering states \cite{McMillan_1968,Furusaki_1991,Kashiwaya_RPP,Herrera_2010,Lu_2015,PhysRevB.92.205424,PhysRevB.96.155426,Lu_2018}, which here involves only local Andreev reflections and normal transmissions, for details see Ref.\,\cite{SM}. To identify the  symmetries of  the pair amplitude $[G^{r}(x,x',\omega)]_{eh}$, we inspect  its dependence on frequency, spin, and spatial coordinates.  The spin symmetry is immediately evident  by writing   $G^{r}_{eh}=\sum_{i}F^{r}_{j}\sigma_{i}$, with $j=0,\cdots,3$. Here,  $F^{r}_{0}$ is the  spin-singlet (S), while $F^{r}_{1,2,3}$ are the equal- ($j=1,2$) and mixed-spin ($j=3$)  triplet (T) states.   Moreover,  we are interested in finding Cooper pair signatures  in  transport observables measured  locally at the interfaces  $x=0, 2L_{\rm S}$. We also know from the previous section that  transport is governed by the topological ABSs that locally reside at $x=L_{\rm S}$. It is therefore most interesting to focus on the local pairs at these interfaces, i.e.~those that are finite for $x=x'$, which means only considering even-parity (E) spatial symmetry. Hence, taking into account that the Cooper pair amplitudes must be overall antisymmetric  due to Fermi-Dirac statistics, we reduce all the possible symmetries to just two local pair symmetries: even-$\omega$ spin-singlet even-parity (ESE) and odd-$\omega$ spin-triplet even-parity (OTE)~\footnote{We note that nonlocal pair amplitudes can emerge if we allow for $x\neq x'$. It is then possible to obtain even-$\omega$ spin-triplet odd-parity and odd-$\omega$ spin-singlet odd-parity pair amplitudes, which are odd under the spatial coordinates \cite{PhysRevB.92.205424,PhysRevB.96.155426}. These components, however, are only finite inside the S regions and thus do not directly contribute to transport.}. 
	Below we discuss the emergence of these two  local pair symmetries in the junction of Fig.\,\ref{Fig1}(a), with  details presented in Ref.\,\cite{SM}.
	
	\begin{figure}[!t]
		\centering
		\includegraphics[width=.48\textwidth]{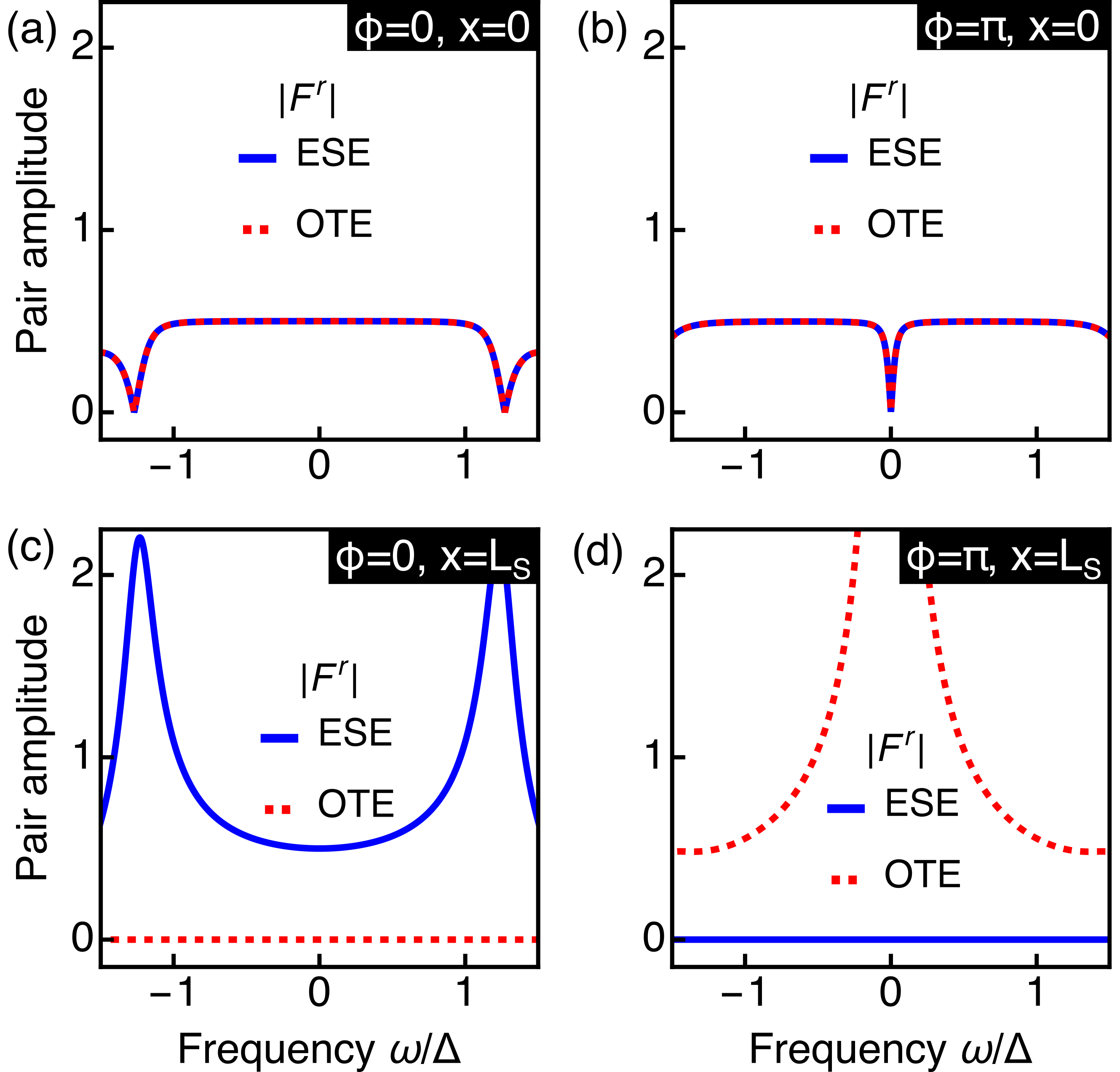} 
		\caption{(a-d) Absolute value of the   local ESE and OTE pair amplitudes inside the topological Josephson junction as a function of $\omega$ at $\phi=0,\pi$ for $x=0$ (a,b) and $x=L_{\rm S}$ (c,d). Parameters same as Fig.~\ref{Fig2}:  $\mu=\Delta$, $L_{\rm S}=2\xi$.}
		\label{Fig3}
	\end{figure} 
	
Under general conditions, we find finite local pair amplitudes with ESE  (OTE) symmetry in all regions of the junction, denoted here as $F^{r, {\rm E(O)}}$, respectively. While the ESE symmetry is in the spin-singlet state  of the parent superconductor, the OTE symmetry is in the mixed spin-triplet state and emerges due to the  strong spin-momentum locking in  2DTI edges \cite{PhysRevB.92.100507,PhysRevB.97.075408,PhysRevB.96.155426,PhysRevResearch.2.022019,Bo_2021}. The existence of these  two pair symmetries in both the N and S regions is determined by the scattering processes at the junction interfaces, which also introduce a strong dependence on the superconducting phase difference $\phi$. 
	At the outer interfaces ($x=0,2L_{\rm S}$), ESE and OTE pairs are simultaneously determined by the same local Andreev reflection, making it  essentially impossible to distinguish between these two types of Cooper pairs. 
	This effect is seen in Fig.\,\ref{Fig3}(a) and Fig.\,\ref{Fig3}(b), which depict the frequency dependence of ESE and OTE amplitudes at $x=0$ for $\phi=0$ and $\pi$, respectively. 
	While at $\phi=0$ both ESE and OTE pairs are finite and constant for $|\omega|<\Delta$, at $\phi=\pi$ both amplitudes develop a dip and vanish at $\omega=0$. 
	
	In the middle of the junction, at $x=L_{\rm S}$, the situation is interestingly distinct to what occurs at the outer interfaces, as seen in Fig.\,\ref{Fig3}(c,d). At $\phi=0$ and $x=L_{\rm S}$ the ESE term is always finite and also captures  the gap edge coherence peaks, while the OTE component completely vanishes. Interestingly, at $\phi=\pi$ and  $x=L_{\rm S}$,  the ESE part  instead vanishes for all $\omega$, while the OTE  term is finite for all $\omega$ and even exhibits a resonant profile at  $\omega=0$, thus becoming both large and the only type of superconducting pairing. 
	The resonant OTE profile results from the protected crossing of the topological ABSs at zero frequency and also reflects the dynamical symmetry of this superconducting state. 
	Indeed, we find that the pair amplitudes at $x=L_{\rm S}$ are fully determined by the scattering processes, which include the ABSs through the zeros of the denominator of the scattering matrix~\cite{Beenakker:92,PhysRevB.96.155426,PhysRevResearch.2.022019}. 
	We have thus identified that, while under general circumstances Cooper pairs with both ESE and OTE symmetries coexist, only  pairs with OTE symmetry remain in the middle of the junction at $\phi=\pi$ and $\omega=0$ \footnote{In the case of a topological Josephson junction with a long middle N region, we have verified that the OTE pairing  remains the only type of pairing in the middle of such N region, also in agreement with previous findings \cite{PhysRevB.96.155426}.  For S regions of distinct lengths, a vanishing small ESE component appears but the OTE pairing remains the largest pair amplitude thus reflecting the dominant character of OTE Cooper pairs, see \cite{SM}.}.
	
	Having established the emergence and dominant behavior of  OTE Cooper pairs at $\phi=\pi$, we finally address their connection to the transport observables discussed in the previous section.  
	Since electron transport at subgap frequencies involves only the transfer of Cooper pairs, the signatures in the Andreev reflection and transmission necessarily correspond to the dominant OTE Cooper pair symmetry. 
	This is also supported by the fact that the origin of the dominant OTE symmetry,  the topological ABSs, is the same as that of the dip and resonant profile in the local Andreev  and transmission probabilities at zero frequency, see Fig.\,\ref{Fig2}(b) and (d), respectively. 
	As a consequence, a clear signature of OTE Cooper pairs is measuring at $\phi=\pi$ either a dip in the local conductance, related to Andreev reflection, or a peak of the nonlocal conductance, determined by normal transmissions.  Due to the topological protection of the zero-energy ABS at $\phi=\pi$, our results remain valid also for longer N regions. In fact, for any N region shorter than the superconducting coherence length, no other states appear in the N region which can modify the results \cite{PhysRevLett.110.017003, PhysRevLett.112.077002, PhysRevB.89.205115, PhysRevB.96.155426}. We have further verified that our results remain overall robust even for longer N regions and also for S regions of distinct lengths \cite{SM}.
	
	In conclusion, we have identified the impact of Cooper pair symmetries in transport observables in a  N-SNS-N topological Josephson junction at the  edge of a 2D topological insulator. In particular,  we have found vanishing  local Andreev reflection  and perfect normal transmission at  $\phi=\pi$ and $\omega=0$ as a result of the interplay between the finite junction size and  the emergence of a pair of topological ABSs. Furthermore, we  have discovered that the  Cooper pairs responsible for this surprising transport behavior only have odd-frequency mixed spin-triplet symmetry in the middle of the junction. 
	We note that very similar  short Josephson junctions at the edge of 2D topological insulators as proposed here have already been fabricated with good proximity-induced superconductivity in, e.g., HgCd/HgTe and InAs/GaSb \cite{Hart_2014,vlad15,Bocquillon_2016,Deacon_2017,Bocquillon17}. In these systems, there is also evidence of having achieved phase tuning to $\phi=\pi$ through the measurement of the $4\pi$ fractional Josephson effect \cite{Bocquillon_2016,Deacon_2017,Bocquillon17}. Moreover, conductance measurements have also been performed in these systems~\cite{Sullivan_2012,Hart_2014,vlad15,Bocquillon17}.    Taken together, these recent results places our proposal clearly within experimental reach.  Our work thus paves the way for highly controllable detection schemes of  odd-frequency pairing in topological  Josephson junctions.  
	
	We acknowledge financial support from the Swedish Research Council  (Vetenskapsr\aa det Grant No.~2021-04121), the G\"{o}ran Gustafsson Foundation, the European Research Council (ERC) under the European Unions Horizon 2020 research and innovation programme (ERC- 2017-StG-757553), and the EU-COST Action CA-16218 Nanocohybri. P. D. acknowledges financial support from Department of Science and Technology (DST), India through SERB Start-up Research Grant (File No. SRG/2022/001121).
		P.~B.~acknowledges support from the Spanish CM ``Talento Program'' Project No. 2019-T1/IND-14088 and the Agencia Estatal de Investigaci\'{o}n Project No. PID2020-117992GA-I00.

	\bibliography{biblio}
\onecolumngrid



\section*{Supplementary material}

# Supplemental Material for "Phase-tunable electron transport assisted by odd-frequency Cooper pairs in topological Josephson junctions"

Jorge Cayao,[1] Paramita Dutta,[2] Pablo Burset,[3] and Annica M. Black-Schaffer[1]

[1]*Department of Physics and Astronomy, Uppsala University, Box 516, S-751 20 Uppsala, Sweden*
[2]*Theoretical Physics Division, Physical Research Laboratory, Navrangpura, Ahmedabad-380009, India*
[3]*Department of Theoretical Condensed Matter Physics,*
*Condensed Matter Physics Center (IFIMAC) and Instituto Nicolás Cabrera,*
*Universidad Autónoma de Madrid, 28049 Madrid, Spain*
(Dated: August 15, 2022)

In this supplementary material we provide the details on the calculation of the scattering amplitudes and the pair amplitudes presented in the main text. We also analyze the effect of junction asymmetry.

## SCATTERING STATES, LOCAL ANDREEV REFLECTION, AND NORMAL TRANSMISSION

The system we study in the main text is a N-SNS-N Josephson junction, with a very short middle N region, formed at the edge of a two-dimensional topological insulator (2DTI) in proximity to conventional spin-singlet $s$-wave superconductors. The interfaces of this system are located at $x = 0, L_{\rm S}, 2L_{\rm S}$, where $L_{\rm S}$ is the length of the S regions. This system is modeled by a Bogoliubov-de Gennes (BdG) Hamiltonian $H_{\rm BdG}$ given by Eq. (1) in the main text. The superconducting order parameter is finite only in S and we assume a finite phase difference $\phi$ between the two S regions. In the main text we discuss equilibrium transport and signatures of induced Cooper pairs. To investigate transport and identify the induced pair amplitudes in our N-SNS-N junction we employ a methodology that involves scattering states, a method that has already proven very powerful in previous works [1–8]. In this section, we provide the details on how we construct the scattering states and then how those allow us to obtain the local Andreev reflection and normal tunneling in leftmost and rightmost N regions, respectively.

### Scattering states

Scattering states account for all the reflections and transmissions at all the interfaces of a junction. In our case, due to the helical nature of the edge states in 2DTIs, there are four different scattering processes at the interfaces of the N-SNS-N junction, which read

$$
\Psi_1(x) = \begin{cases} \phi_1^{\rm N}\,{\rm e}^{ik_e x} + a_1\phi_3^{\rm N}\,{\rm e}^{ik_h x} + b_1\phi_2^{\rm N}\,{\rm e}^{-ik_e x}, & x < 0 \\ p_1\phi_1^{\rm S}\,{\rm e}^{ik_e^{\rm S}x} + q_1\phi_2^{\rm S}\,{\rm e}^{-ik_e^{\rm S}x} + r_1\phi_3^{\rm S}\,{\rm e}^{ik_h^{\rm S}x} + s_1\phi_4^{\rm S}\,{\rm e}^{-ik_h^{\rm S}x}, & 0 < x < L_{\rm S} \\ h_1\phi_1^{\rm S}\,{\rm e}^{ik_e^{\rm S}x} + e_1\phi_2^{\rm S}\,{\rm e}^{-ik_e^{\rm S}x} + f_1\phi_3^{\rm S}\,{\rm e}^{ik_h^{\rm S}x} + g_1\phi_4^{\rm S}\,{\rm e}^{-ik_h^{\rm S}x}, & L_{\rm S} < x < 2L_{\rm S} \\ c_1\phi_1^{\rm N}\,{\rm e}^{ik_e x} + d_1\phi_4^{\rm N}\,{\rm e}^{-ik_h x}, & x > 2L_{\rm S} \end{cases},
$$

$$
\Psi_2(x) = \begin{cases} \phi_4^{\rm N}\,{\rm e}^{-ik_h x} + a_2\phi_2^{\rm N}\,{\rm e}^{-ik_e x} + b_2\phi_3^{\rm N}\,{\rm e}^{ik_h x}, & x < 0 \\ p_2\phi_1^{\rm S}\,{\rm e}^{ik_e^{\rm S}x} + q_2\phi_2^{\rm S}\,{\rm e}^{-ik_e^{\rm S}x} + r_2\phi_3^{\rm S}\,{\rm e}^{ik_h^{\rm S}x} + s_2\phi_4^{\rm S}\,{\rm e}^{-ik_h^{\rm S}x}, & 0 < x < L_{\rm S} \\ h_2\phi_1^{\rm S}\,{\rm e}^{ik_e^{\rm S}x} + e_2\phi_2^{\rm S}\,{\rm e}^{-ik_e^{\rm S}x} + f_2\phi_3^{\rm S}\,{\rm e}^{ik_h^{\rm S}x} + g_2\phi_4^{\rm S}\,{\rm e}^{-ik_h^{\rm S}x}, & L_{\rm S} < x < 2L_{\rm S} \\ c_2\phi_4^{\rm N}\,{\rm e}^{-ik_h x} + d_2\phi_1^{\rm N}\,{\rm e}^{ik_e x}, & x > 2L_{\rm S} \end{cases},
$$

$$
\Psi_3(x) = \begin{cases} c_3\phi_2^{\rm N}\,{\rm e}^{-ik_e x} + d_3\phi_3^{\rm N}\,{\rm e}^{ik_h x}, & x < 0 \\ p_3\phi_1^{\rm S}\,{\rm e}^{ik_e^{\rm S}x} + q_3\phi_2^{\rm S}\,{\rm e}^{-ik_e^{\rm S}x} + r_3\phi_3^{\rm S}\,{\rm e}^{ik_h^{\rm S}x} + s_3\phi_4^{\rm S}\,{\rm e}^{-ik_h^{\rm S}x}, & 0 < x < L_{\rm S} \\ h_3\phi_1^{\rm S}\,{\rm e}^{ik_e^{\rm S}x} + e_3\phi_2^{\rm S}\,{\rm e}^{-ik_e^{\rm S}x} + f_3\phi_3^{\rm S}\,{\rm e}^{ik_h^{\rm S}x} + g_3\phi_4^{\rm S}\,{\rm e}^{-ik_h^{\rm S}x}, & L_{\rm S} < x < 2L_{\rm S} \\ \phi_2^{\rm N}\,{\rm e}^{-ik_e x} + a_3\phi_4^{\rm N}\,{\rm e}^{-ik_h x} + b_3\phi_1^{\rm N}\,{\rm e}^{ik_e x}, & x > 2L_{\rm S} \end{cases},
$$

$$
\Psi_4(x) = \begin{cases} c_4\phi_3^{\rm N}\,{\rm e}^{ik_h x} + d_4\phi_2^{\rm N}\,{\rm e}^{-ik_e x}, & x < 0 \\ p_4\phi_1^{\rm S}\,{\rm e}^{ik_e^{\rm S}x} + q_4\phi_2^{\rm S}\,{\rm e}^{-ik_e^{\rm S}x} + r_4\phi_3^{\rm S}\,{\rm e}^{ik_h^{\rm S}x} + s_4\phi_4^{\rm S}\,{\rm e}^{-ik_h^{\rm S}x}, & 0 < x < L_{\rm S} \\ h_4\phi_1^{\rm S}\,{\rm e}^{ik_e^{\rm S}x} + e_4\phi_2^{\rm S}\,{\rm e}^{-ik_e^{\rm S}x} + f_4\phi_3^{\rm S}\,{\rm e}^{ik_h^{\rm S}x} + g_4\phi_4^{\rm S}\,{\rm e}^{-ik_h^{\rm S}x}, & L_{\rm S} < x < 2L_{\rm S} \\ \phi_3^{\rm N}\,{\rm e}^{ik_h x} + a_4\phi_1^{\rm N}\,{\rm e}^{ik_e x} + b_4\phi_4^{\rm N}\,{\rm e}^{-ik_h x}, & x > 2L_{\rm S} \end{cases},
$$

(S1)

where

$$\phi_1^{\mathrm{N}} = \big(1,0,0,0\big)^T, \quad \phi_1^{\mathrm{S}} = \big(u,0,v,0\big)^T,$$
$$\phi_2^{\mathrm{N}} = \big(0,1,0,0\big)^T, \quad \phi_2^{\mathrm{S}} = \big(0,u,0,v\big)^T,$$
$$\phi_3^{\mathrm{N}} = \big(0,0,1,0\big)^T, \quad \phi_3^{\mathrm{S}} = \big(v,0,u,0\big)^T,$$
$$\phi_4^{\mathrm{N}} = \big(0,0,0,1\big)^T, \quad \phi_4^{\mathrm{S}} = \big(0,v,0,u\big)^T,$$

$$(\mathrm{S2})$$

are wavefunctions of $H_{\mathrm{BdG}}(k)$ (the Fourier transform of Eq. (1) in the main text) and

$$k_{e,h} = (\mu \pm \omega)/v_F,$$
$$k_{e,h}^{\mathrm{S}} = (\mu \pm \sqrt{\omega^2 - \Delta^2})/v_{\mathrm{F}} = k_\mu \pm k(\omega),$$
$$u,v = \sqrt{\frac{1}{2}\left[1 \pm \frac{\sqrt{\omega^2 - \Delta^2}}{\omega}\right]},$$

$$(\mathrm{S3})$$

with $k_{\mathrm{F}} = \mu/v_{\mathrm{F}}$ and $k = \sqrt{\omega^2 - \Delta^2}/v_{\mathrm{F}}$.

As mentioned above, the scattering states $\Psi_i$ account for the reflections and transmissions in the junction. Hence, $\Psi_1$ has a clear physical interpretation. For instance, $\Psi_1$ represents that an incident electron from the leftmost N region (with wavevector $k_e$ and wavefunction $\phi_1^{\mathrm{N}}$) can be reflected as an electron or hole at $x = 0$ with probability amplitudes $a_1$ and $b_1$, respectively. It can be also transmitted into electron and hole states in the S regions, where reflections and transmissions occur at each interface until it is transmitted to the rightmost N region with amplitudes $c_1$ and $d_1$ for electron and hole states, respectively. The reflection and transmission processes in the S regions occur due to the finite length of the S regions and the remaining coefficients in $\Psi_1$ also have a clear meaning: the amplitudes $p_1$ and $s_1$ characterize the electron and hole transmission states in the left S region, while the amplitudes $q_1$ and $r_1$ the electron and hole reflections; similarly for the right S region and amplitudes $h_1, g_1$ and $e_1, f_1$. An analogous interpretation also applies for $\Psi_{2,3,4}$. Before going further, it is also important to mention that the reflection of an electron (hole) into a hole (electron) in the same region is known as Andreev reflection or, more accurately, local Andreev reflection, while electron (hole) into an electron (hole) is known as normal reflection. Similarly, the transmission of an electron (hole) state into an electron (hole) state is known as normal transmission, while transmission of an electron (hole) state into an hole (electron) is known as Andreev transmission, nonlocal Andreev reflection, or crossed Andreev reflection.

In order to fully characterize the scattering states given by Eqs. (S1), we need to find all the coefficients $a_i, b_i, c_i, d_i, p_i, q_i, r_i, s_i, h_i, e_i, f_i, g_i$. These coefficients can be obtained by matching $\Psi_i$ at each of the interfaces of the N-SNS-N junction, namely at $x = 0$, $x = L_{\mathrm{S}}$, and at $x = 2L_{\mathrm{S}}$,

$$\left[\Psi_i(x<0)\right]_{x=0} = \left[\Psi_i(0<x<L_{\mathrm{S}})\right]_{x=0},$$
$$\left[\Psi_i(0<x<L_{\mathrm{S}})\right]_{x=L_{\mathrm{S}}} = \left[\Psi_i(L_{\mathrm{S}}<x<2L_{\mathrm{S}})\right]_{x=L_{\mathrm{S}}},$$
$$\left[\Psi_i(L_{\mathrm{S}}<x<2L_{\mathrm{S}})\right]_{x=2L_{\mathrm{S}}} = \left[\Psi_i(x>2L_{\mathrm{S}})\right]_{x=2L_{\mathrm{S}}}.$$

$$(\mathrm{S4})$$

We stress that each scattering state $\Psi_i$ is a four column vector, which then provides four equations for each $i = 1, 2, 3, 4$. Based on Eqs. (S4), in the end, we thus have a system of 48 equations that allows us to find the 48 unknown coefficients of the scattering states $\Psi_i$. Furthermore, we point out that in Eqs. (S1) we have written the scattering wavefunctions in a general form. However, normal reflection and non-local Andreev transmission are forbidden by helicity conservation of the 2DTI edge [9, 10]: an incident electron can be only reflected as a hole by a superconducting barrier or transmitted as an electron through it. That is why, after matching the scattering states above, we obtain

$$b_i = 0, \quad d_i = 0, \quad p_{2,3} = 0, \quad q_{1,4} = 0, \quad r_{2,3} = 0,$$
$$s_{1,4} = 0, \quad e_{1,4} = 0, \quad f_{2,3} = 0, \quad g_{1,4} = 0, \quad h_{2,3} = 0.$$

$$(\mathrm{S5})$$

For the rest of the coefficients we obtain expressions that depend on the system parameters, including the superconducting phase difference $\phi$ and the length of the S regions. For instance, the expressions for the Andreev reflections

$a_{1,2}$ and normal transmissions $c_{1,2}$ are given by

$$a_1 = \frac{\left[\mathrm{e}^{ik_e^{\mathrm{S}}L_{\mathrm{S}}} - \mathrm{e}^{ik_h^{\mathrm{S}}L_{\mathrm{S}}}\right]\left[\frac{u}{v}\left(\mathrm{e}^{ik_h^{\mathrm{S}}L_{\mathrm{S}}} + \mathrm{e}^{i(k_e^{\mathrm{S}}L_{\mathrm{S}} - \phi)}\right) - \frac{v}{u}\left(\mathrm{e}^{ik_e^{\mathrm{S}}L_{\mathrm{S}}} + \mathrm{e}^{i(k_h^{\mathrm{S}}L_{\mathrm{S}} - \phi)}\right)\right]}{\mathrm{e}^{-i\phi}\left[\mathrm{e}^{ik_e^{\mathrm{S}}L_{\mathrm{S}}} - \mathrm{e}^{ik_h^{\mathrm{S}}L_{\mathrm{S}}}\right]^2 - \left[\frac{u}{v}\mathrm{e}^{ik_h^{\mathrm{S}}L_{\mathrm{S}}} - \frac{v}{u}\mathrm{e}^{ik_e^{\mathrm{S}}L_{\mathrm{S}}}\right]^2},$$

$$a_2 = \frac{\left[\mathrm{e}^{ik_e^{\mathrm{S}}L_{\mathrm{S}}} - \mathrm{e}^{ik_h^{\mathrm{S}}L_{\mathrm{S}}}\right]\left[\frac{u}{v}\left(\mathrm{e}^{ik_h^{\mathrm{S}}L_{\mathrm{S}}} + \mathrm{e}^{i(k_e^{\mathrm{S}}L_{\mathrm{S}} + \phi)}\right) - \frac{v}{u}\left(\mathrm{e}^{ik_e^{\mathrm{S}}L_{\mathrm{S}}} + \mathrm{e}^{i(k_h^{\mathrm{S}}L_{\mathrm{S}} + \phi)}\right)\right]}{\mathrm{e}^{i\phi}\left[\mathrm{e}^{ik_e^{\mathrm{S}}L_{\mathrm{S}}} - \mathrm{e}^{ik_h^{\mathrm{S}}L_{\mathrm{S}}}\right]^2 - \left[\frac{u}{v}\mathrm{e}^{ik_h^{\mathrm{S}}L_{\mathrm{S}}} - \frac{v}{u}\mathrm{e}^{ik_e^{\mathrm{S}}L_{\mathrm{S}}}\right]^2},$$

$$c_1 = \frac{\left[\frac{u}{v} - \frac{v}{u}\right]^2 \mathrm{e}^{-2i(k_e - k_e^{\mathrm{S}} - k_h^{\mathrm{S}})L_{\mathrm{S}}}}{-\mathrm{e}^{-i\phi}\left[\mathrm{e}^{ik_e^{\mathrm{S}}L_{\mathrm{S}}} - \mathrm{e}^{ik_h^{\mathrm{S}}L_{\mathrm{S}}}\right]^2 + \left[\frac{u}{v}\mathrm{e}^{ik_h^{\mathrm{S}}L_{\mathrm{S}}} - \frac{v}{u}\mathrm{e}^{ik_e^{\mathrm{S}}L_{\mathrm{S}}}\right]^2},$$

$$c_2 = \frac{\left[\frac{u}{v} - \frac{v}{u}\right]^2 \mathrm{e}^{2ik_h L_{\mathrm{S}}}}{-\mathrm{e}^{i\phi}\left[\mathrm{e}^{ik_e^{\mathrm{S}}L_{\mathrm{S}}} - \mathrm{e}^{ik_h^{\mathrm{S}}L_{\mathrm{S}}}\right]^2 + \left[\frac{u}{v}\mathrm{e}^{ik_h^{\mathrm{S}}L_{\mathrm{S}}} - \frac{v}{u}\mathrm{e}^{ik_e^{\mathrm{S}}L_{\mathrm{S}}}\right]^2}.$$

(S6)

The rest of the coefficients acquire a more complicated form but they are also not necessary to characterize equilibrium electron transport in our setup, so we do not report them here.

In the expressions given by Eq. (S6), $a_{1(2)}$ represents the Andreev reflection amplitude of an incident electron (hole) with spin up in the leftmost N region into a hole (electron) with spin down in the same leftmost N region. This process allows the transfer of a Cooper pair into the S region. On the other hand, $c_{1(2)}$ represent the tunneling amplitude of an incident electron (hole) in the leftmost N region into an electron (hole) in the rightmost N region. In the main text we have used the labels $r_{eh(he)}$ and $t_{ee(hh)}$ instead of $a_{1(2)}$ and $c_{1(2)}$, respectively, to highlight the physical relevance of these coefficients. By a simple inspection is not complicated to observe that $r_{he}(\phi) = r_{eh}(-\phi)$ and $t_{hh}(\phi) = t_{ee}(-\phi)$, relations used in the main text to evaluate the probabilities of the total Andreev reflection $R = |r_{eh}|^2 + |r_{he}|^2$ and total normal transmission $T = |t_{ee}|^2 + |t_{hh}|^2$. Taken together, Eqs. (S6) determine the local Andreev reflection and normal transmission in the leftmost and rightmost N regions, respectively.

An interesting property of the Andreev and transmission amplitudes in our setup is that, due to the strong phase dependence, they capture the emergence of the Andreev bound states at the junction between the two superconductors. This is possible because the zeroes in the denominator of these scattering amplitudes actually corresponds to the poles of the scattering matrix, which gives the conditions for the formation of bound states in a scattering system, see, e.g., Refs. [7, 11, 12]. Following this reasoning, in our case, $r_{eh(he)}$ and $t_{ee(hh)}$ capture the emergence of the Andreev bound states, which is discussed in the main text. In particular, in Fig. 2 in the main text we show that the probabilities $R$ and $T$ directly reflect the emergence of a pair of Andreev bound states, including the development of a topological protected crossing at $\phi = \pi$ and $\omega = 0$. Moreover, the behavior of $R$ and $T$ in this regime can be seen by obtaining $r_{eh}$ and $t_{ee}$ at $\omega = 0$, where we obtain

$$r_{eh}(\omega = 0) = \frac{1}{2i}\frac{\sinh(2\kappa L_{\mathrm{S}})(1 + \mathrm{e}^{-i\phi})}{\sinh^2(\kappa L_{\mathrm{S}})\mathrm{e}^{-i\phi} + \cosh^2(\kappa L_{\mathrm{S}})},$$

$$t_{ee}(\omega = 0) = \frac{1}{\sinh^2(\kappa L_{\mathrm{S}})\mathrm{e}^{-i\phi} + \cosh^2(\kappa L_{\mathrm{S}})},$$

(S7)

where $\kappa = 1/\xi$, with $\xi = v_{\mathrm{F}}/\Delta$ being the superconducting coherence length, and $v_{\mathrm{F}}$ the Fermi velocity. From Eqs. (S7) it is evident that $r_{eh} = 0$ and $t_{ee} = 1$ at $\phi = \pi$ when $2\kappa L_{\mathrm{S}} > 1$. This is discussed in the main text, where Eq. (S7) is reproduced as Eqs. (2).

### Impact of S region asymmetry on the Andreev and transmission probabilities

In the discussion above we considered superconducting regions of the same length $L_{\mathrm{S}}$. Under realistic experimental conditions it might be challenging to obtain S segments of exactly the same length, although several experiments have shown that it might be a quite feasible task in the near future, see e.g., Refs. [13–17]. To account for this potential issue, we here discuss the impact of having S segments of distinct lengths. For this purpose, we consider a N-SNS-N junction with interfaces located at $x = 0$, $x = L_{\mathrm{L}}$, and $x = L_{\mathrm{L}} + L_{\mathrm{R}}$, where $L_{\mathrm{L(R)}}$ represent the length of the left (right) S region.

The scattering states $\Psi_i$ in this case are again determined by Eqs. (S1) with the fourth line now corresponding to the right S regions, now with $L_{\mathrm{L}} < x < L_{\mathrm{L}} + L_{\mathrm{R}}$. To find the scattering state coefficients we proceed as before and match $\Psi_i$ at each of the interfaces of the N-SNS-N junction, namely at $x = 0$, $x = L_{\mathrm{L}}$, and $x = L_{\mathrm{L}} + L_{\mathrm{R}}$. Then,

Eqs. (S4) is slightly modified and become

$$
\begin{aligned}
\left[\Psi_i(x<0)\right]_{x=0} &= \left[\Psi_i(0<x<L_{\mathrm{L}})\right]_{x=0}, \\
\left[\Psi_i(0<x<L_{\mathrm{L}})\right]_{x=L_{\mathrm{L}}} &= \left[\Psi_i(L_{\mathrm{L}}<x<L_{\mathrm{L}}+L_{\mathrm{R}})\right]_{x=L_{\mathrm{L}}}, \\
\left[\Psi_i(L_{\mathrm{L}}<x<L_{\mathrm{L}}+L_{\mathrm{R}})\right]_{x=L_{\mathrm{L}}+L_{\mathrm{R}}} &= \left[\Psi_i(x>L_{\mathrm{L}}+L_{\mathrm{R}})\right]_{x=L_{\mathrm{L}}+L_{\mathrm{R}}}.
\end{aligned}
\tag{S8}
$$

By using Eqs. (S8), we obtain all scattering coefficients and can characterize the scattering states. We find that Eqs. (S5) remain, in agreement with expectation as they stem from the helical nature of the system. The Andreev reflections and transmissions are however modified due to the distinct lengths of S and are now given by

$$
\begin{aligned}
a_1 &= \frac{uv\left\{\left[e^{i(k_e^S L_{\mathrm{L}}+k_e^S L)}-e^{i(k_e^S L_{\mathrm{L}}+k_h^S L)}\right]\left[v^2 e^{ik_h^S L_{\mathrm{L}}}-u^2 e^{ik_e^S L_{\mathrm{L}}}\right]-Pe^{i\phi}\left[u^2 e^{i(k_e^S L_{\mathrm{L}}+k_h^S L)}-v^2 e^{i(k_h^S L_{\mathrm{L}}+k_e^S L)}\right]\right\}}{P\left[e^{i(k_e^S L_{\mathrm{L}}+k_h^S L)}-e^{i(k_h^S L_{\mathrm{L}}+k_e^S L)}\right]u^2 v^2 + e^{i\phi}\left[u^2 e^{ik_h^S L_{\mathrm{L}}}-v^2 e^{ik_e^S L_{\mathrm{L}}}\right]\left[u^2 e^{i(k_e^S L_{\mathrm{L}}+k_h^S L)}-v^2 e^{i(k_h^S L_{\mathrm{L}}+k_e^S L)}\right]}, \\
a_2 &= \frac{uv\left\{e^{i\phi}\left[e^{i(k_h^S L_{\mathrm{L}}+k_e^S L)}-e^{i(k_e^S L_{\mathrm{L}}+k_h^S L)}\right]\left[v^2 e^{ik_e^S L_{\mathrm{L}}}-u^2 e^{ik_e^S L_{\mathrm{L}}}\right]-P\left[u^2 e^{i(k_e^S L_{\mathrm{L}}+k_h^S L)}-v^2 e^{i(k_h^S L_{\mathrm{L}}+k_e^S L)}\right]\right\}}{Pe^{i\phi}\left[e^{i(k_e^S L_{\mathrm{L}}+k_h^S L)}-e^{i(k_h^S L_{\mathrm{L}}+k_e^S L)}\right]u^2 v^2 + \left[u^2 e^{ik_h^S L_{\mathrm{L}}}-v^2 e^{ik_e^S L_{\mathrm{L}}}\right]\left[u^2 e^{i(k_e^S L_{\mathrm{L}}+k_h^S L)}-v^2 e^{i(k_h^S L_{\mathrm{L}}+k_e^S L)}\right]}, \\
c_1 &= \frac{(u^2-v^2)^2 e^{i\phi}e^{-ik_e L}e^{i(k_e^S+k_h^S)(L_{\mathrm{L}}+L)}}{P\left[e^{i(k_e^S L_{\mathrm{L}}+k_h^S L)}-e^{i(k_h^S L_{\mathrm{L}}+k_e^S L)}\right]u^2 v^2 + e^{i\phi}\left[u^2 e^{ik_h^S L_{\mathrm{L}}}-v^2 e^{ik_e^S L_{\mathrm{L}}}\right]\left[u^2 e^{i(k_e^S L_{\mathrm{L}}+k_h^S L)}-v^2 e^{i(k_h^S L_{\mathrm{L}}+k_e^S L)}\right]}, \\
c_2 &= \frac{(u^2-v^2)^2 e^{ik_h L}e^{i(k_e^S+k_h^S)L_{\mathrm{L}}}}{Pe^{i\phi}\left[e^{i(k_e^S L_{\mathrm{L}}+k_h^S L)}-e^{i(k_h^S L_{\mathrm{L}}+k_e^S L)}\right]u^2 v^2 + \left[u^2 e^{ik_h^S L_{\mathrm{L}}}-v^2 e^{ik_e^S L_{\mathrm{L}}}\right]\left[u^2 e^{i(k_e^S L_{\mathrm{L}}+k_h^S L)}-v^2 e^{i(k_h^S L_{\mathrm{L}}+k_e^S L)}\right]},
\end{aligned}
\tag{S9}
$$

where $L=L_{\mathrm{L}}+L_{\mathrm{R}}$ and $P=\left(e^{ik_e^S L_{\mathrm{L}}}-e^{ik_h^S L_{\mathrm{L}}}\right)$. At zero frequency, the previous expressions reduce to

$$
\begin{aligned}
a_1(\omega=0) &= \frac{1}{i}\frac{\sinh(L_{\mathrm{L}}\kappa)\cosh(L_{\mathrm{R}}\kappa)+e^{-i\phi}\sinh(L_{\mathrm{R}}\kappa)\cosh(L_{\mathrm{L}}\kappa)}{\sinh(L_{\mathrm{L}}\kappa)\sinh(L_{\mathrm{R}}\kappa)e^{-i\phi}+\cosh(L_{\mathrm{L}}\kappa)\cosh(L_{\mathrm{R}}\kappa)}, \\
c_1(\omega=0) &= \frac{1}{\sinh(L_{\mathrm{L}}\kappa)\sinh(L_{\mathrm{R}}\kappa)e^{-i\phi}+\cosh(L_{\mathrm{L}}\kappa)\cosh(L_{\mathrm{R}}\kappa)},
\end{aligned}
\tag{S10}
$$

which become Eqs. (S7) when $L_{\mathrm{L}}=L_{\mathrm{R}}\equiv L_{\mathrm{S}}$, as they should.

To visualize the impact of the distinct length of the S segments, we plot in Fig. S1(a,e) the total local Andreev reflection $|R|^2=|r_{eh}|^2+|r_{he}|^2$ and the total transmission $|T|^2=|t_{ee}|^2+|t_{hh}|^2$, respectively, as a function of frequency $\omega$ and phase difference $\phi$ for $L_{\mathrm{L}}=2\xi$ and $L_{\mathrm{R}}=1.8\xi$. In Fig. S1(b-d,f-h) we additionally show $|r_{eh}|^2$ and $|t_{ee}|^2$ as a function of $\omega$ at $\phi=\pi$, length of the right S region $L_{\mathrm{R}}$ at $\phi=\pi$ and $\omega=0$, and superconducting phase difference at $\omega=0$, respectively. An immediate observation is that the total Andreev and transmission probabilities still acquire a strong phase dependence for $|\omega|<\Delta$, as in the case where the S segments have the same length, see Fig. 2 in the main text. It is thus evident that these scattering probabilities capture the formation of the pair of topological ABSs at $x=L_{\mathrm{L}}$ even when the S regions have distinct lengths. We have verified that this feature is preserved even by further decreasing/increasing the asymmetry in the lengths of the S regions. Even though the strong phase-dependent regions in the total probabilities reveal the formation of topological ABSs, there exists some effect due to the distinct lengths, which is observed in Fig. S1(b-d,f-h). However, whenever the variations in the lengths are not too large, e.g. for all lengths used in Fig. S1(a,b,d,e,f,h), as would be expected for junctions designed to be symmetric, the Andreev (transmission) probabilities still exhibit a dip and peak at $\omega=0$ and $\phi=\pi$, albeit not exactly reaching the vanishing (perfect) value discussed in the main text. We have verified that the dip (peak) profile in the Andreev (transmission) probability is seen even when the asymmetry is as large as 50%, which highlights the robustness of the transport features that in the end permit a clear identification of the topological ABSs and odd-frequency pair amplitude.

## RETARDED GREEN'S FUNCTIONS AND SUPERCONDUCTING PAIRING

In this section we outline the method we employ to obtain the superconducting pair amplitudes in our N-SNS-N Josephson junction. We follow the formalism based on retarded and advanced Green's functions, corresponding to outgoing and incoming waves, respectively. In general, the retarded Green's function can be calculated using [1–8, 18]

$$
G^r(x,x',\omega)=\begin{cases}\alpha_1\Psi_1(x)\tilde{\Psi}_3^T(x')+\alpha_2\Psi_1(x)\tilde{\Psi}_4^T(x')+\alpha_3\Psi_2(x)\tilde{\Psi}_3^T(x')+\alpha_4\Psi_2(x)\tilde{\Psi}_4^T(x'), & x>x', \\ \beta_1\Psi_3(x)\tilde{\Psi}_1^T(x')+\beta_2\Psi_4(x)\tilde{\Psi}_1^T(x')+\beta_3\Psi_3(x)\tilde{\Psi}_2^T(x')+\beta_4\Psi_4(x)\tilde{\Psi}_2^T(x'), & x<x',\end{cases}
\tag{S11}
$$

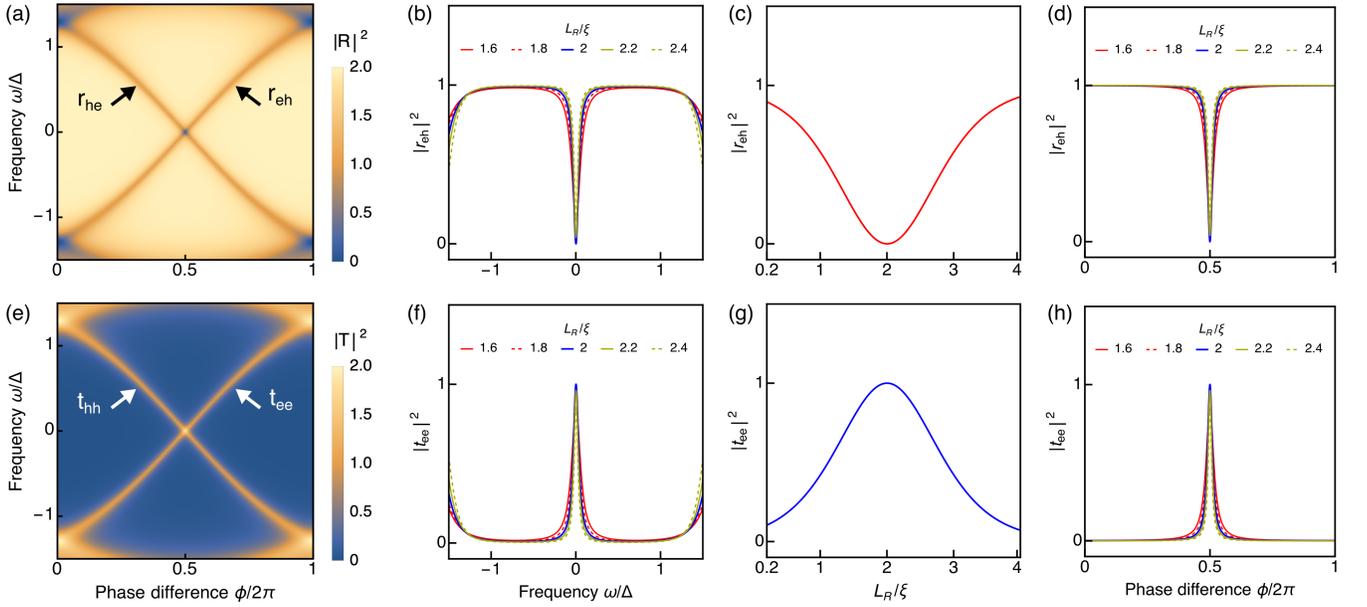

FIG. S1. Scattering probabilities for asymmetric junctions. (a) Total Andreev reflection probability $|R|^2 = |r_{eh}|^2 + |r_{he}|^2$ due to the incidence of an electron ($r_{eh}$) and a hole ($r_{he}$) from the leftmost N region as a function of $\omega$ and $\phi$ for $L_L = 2\xi$ and $L_R = 1.8\xi$. Black arrows indicate the phase-dependent branches from $r_{eh(he)}$. (b-d) Andreev probability $|r_{eh}|^2$ as a function of frequency at $\phi = \pi$ (b), length of right S region $L_R$ at $\phi = \pi$ and $\omega = 0$ (c), and superconducting phase difference at $\omega = 0$ (d).(e-h) Same as (a-d) but for the transmission $|T|^2 = |t_{ee}|^2 + |t_{hh}|^2$ to the rightmost N region. Parameters: $\mu = v_F = \Delta = 1$, $L_L = 2\xi$.

where $\Psi_i$, $i = 1, 2, 3, 4$ are the scattering states given by Eqs. (S1), while $\tilde{\Psi}_i$ correspond to the conjugated processes obtained using $\tilde{H}_{\text{BdG}}(k) = H_{\text{BdG}}^*(-k) = H_{\text{BdG}}^T(-k)$ instead of Eq. (1) in the main text. The structure of these conjugated processes is similar to the normal processes given by Eqs. (S1), but with distinct coefficients which are also obtained following Eqs. (S4).

The retarded Green's function satisfies the equation of motion

$$[\omega - H_{BdG}(x)]G^r(x, x', \omega) = \delta(x - x'),$$
(S12)

where its integration around $x = x'$,

$$\left[G^r(x > x')\right]_{x=x'} - \left[G^r(x < x')\right]_{x=x'} = \frac{\sigma_z \tau_z}{i v_F},$$
(S13)

provides a system of equations that allows us to find the coefficients $\alpha_i$ and $\beta_i$ in Eq. (S11). Once the coefficients $\alpha_i$ and $\beta_i$ are found, using Eq. (S13), and the retarded Green's functions are fully characterized, the advanced Green's function can be directly calculated using $G^a(x, x', \omega) = [G^r(x', x, \omega)]^\dagger$. The Green's functions $G^{r(a)}$ are $4 \times 4$ matrices in Nambu space,

$$G^r(x, x', \omega) = \begin{pmatrix} G^r_{ee}(x, x', \omega) & G^r_{eh}(x, x', \omega) \\ G^r_{he}(x, x', \omega) & G^r_{hh}(x, x', \omega) \end{pmatrix},$$
(S14)

where the diagonal elements are the regular electron-electron and hole-hole Green's functions, while the off-diagonal elements corresponds to the anomalous, i.e. superconducting, electron-hole part. Here each element of $G^r$ is a $2 \times 2$ matrix in spin-space and is directly connected to observables. For instance, the electron-electron part allows the calculation of the local density of states, while the anomalous electron-hole term is the superconducting pair amplitudes. Here we focus primarily on the latter.

Once the retarded and advanced Green's functions are calculated, we can decompose the spin symmetry of the anomalous electron-hole part as

$$G^{r(a)}_{eh}(x, x', \omega) = F_0^{r(a)}(x, x', \omega)\sigma_0 + \sum_{j=1}^{3} F_j^{r(a)}(x, x', \omega)\sigma_j,$$
(S15)

where $F_0^{r(a)}$ corresponds to spin-singlet, $F_{1,2}^{r(a)}$ to equal spin-triplet and $F_3^{r(a)}$ to mixed spin-triplet retarded (advanced) pair amplitudes. In order to disentangle the exact symmetry of the pair amplitude $F_j^r$, it is also necessary to decompose these singlet (S) and triplet (T) components into their even (E) and odd (O) dependencies in both space and frequency. In general, the combination of frequency, spin, and spatial coordinates lead to four pair symmetry classes: even-$\omega$ spin-singlet even-parity (ESE), even-$\omega$ spin-triplet odd-parity (ETO), odd-$\omega$ spin-singlet odd-parity (OSO), and odd-$\omega$ spin-triplet even-parity (OTE). While all these four classes are allowed, we here only focus on the pair amplitudes that are local in space, namely, being finite at $x = x'$, thus leaving only the ESE and OTE amplitudes. This is motivated by the fact that it is these local pair amplitudes that allow us to make contact with the conditions at which transport is measured. In fact, transport is commonly measured locally at the leftmost ($x = x' = 0$) and rightmost ($x = x' = 2L_S$) interfaces. Furthermore, the formation of the pair of topological Andreev bound states in the middle of the junction occurs locally at $x = x' = L_S$, where the S regions exhibit a change in their superconducting phases. This discussion of local pair amplitudes supports the results in the section "Induced odd-$\omega$ pairs" in the main text. In what follows we employ the scattering states obtained in Eqs. (S1) to calculate the Green's function following Eq. (S11) in both the outer N and S regions of the N-SNS-N junctions.

### Pair amplitude in the outer N regions

We first obtain the Green's function in the left- and rightmost N regions. Given the symmetry of the junction, it is enough to find the Green's function in the leftmost region. For notation purposes, we label the left (right) N regions as $N_{L(R)}$. After some algebra, we obtain that the electron-hole component in the leftmost N region is given by

$$[G_{N_L}^r(x, x', \omega)]_{eh} = \frac{1}{iv_F} \begin{pmatrix} 0 & 0 \\ 0 & a_2 e^{-i(k_e x - k_h x')} \end{pmatrix} \tag{S16}$$

which can be decomposed using Eq. (S15) to obtain

$$\begin{aligned} F_{0,N_L}^r(x, x') &= \frac{a_2}{2iv_F} e^{-i(k_e x - k_h x')}, \\ F_{3,N_L}^r(x, x') &= -\frac{a_2}{2iv_F} e^{-i(k_e x - k_h x')}, \end{aligned} \tag{S17}$$

which correspond to the spin-singlet and mixed-spin triplet pair amplitudes, respectively. Under general conditions, both of these pair amplitudes exhibit even- and odd-dependencies under the exchange of spatial coordinates. However, as discussed in the beginning of this section, we only focus on the local pair amplitudes, namely, those finite for $x = x'$, which allows us to obtain the ESE and OTE pair symmetries given by

$$\begin{aligned} F_{N_L}^{r,E}(x) &= \frac{a_2}{2iv_F} e^{-i(k_e - k_h)x}, \\ F_{N_L}^{r,O}(x) &= -\frac{a_2}{2iv_F} e^{-i(k_e - k_h)x}. \end{aligned} \tag{S18}$$

As seen, both of these are determined by the Andreev reflection amplitude $a_2$, denoted as $r_{he}$ in the main text. Thus, if the Andreev reflection is zero (finite), then the ESE and OTE pair amplitudes are zero (finite) in $N_L$, indicating that it is impossible to distinguish between these two induced pair symmetries as both are simultaneously determined by the same Andreev reflection coefficient. In the rightmost N region, the pair amplitudes have similar expressions as in Eqs. (S18), but with the Andreev reflection in that region instead of $a_2$. These results of the pair amplitudes in the outer N regions, given by Eq. (S18), are reported in the section "Induced odd-$\omega$ pairs" of the main text, where we for simplicity drop the subscript L.

### Pair amplitude in the S regions

We next derive the pair amplitudes in the S regions, which are obtained after calculating the retarded Green's functions in the S regions following the same method as in the previous subsection. Due to the symmetry of the junction, it is again sufficient to only obtain the Green's function in the left S region. For the electron-hole component in the left S, we obtain

$$[G_{S_L}^r(x, x')]_{eh} = \begin{pmatrix} [G_{S_L}^r(x, x')]_{eh, \uparrow\downarrow} & 0 \\ 0 & [G_{S_L}^r(x, x')]_{eh, \downarrow\uparrow} \end{pmatrix}, \tag{S19}$$

$$
\begin{aligned}
\left[G^r_{\mathrm{S_L}}(x,x')\right]_{eh,\uparrow\downarrow} = {}& u^2 \mathrm{e}^{i(k^{\mathrm{S}}_e x - k^{\mathrm{S}}_h x')}\left[\theta(x-x')p_1\tilde{s}_3\alpha_1 + \theta(x'-x)p_4\tilde{s}_2\beta_4\right] \\
& + uv\mathrm{e}^{i(x-x')k^{\mathrm{S}}_e}\left[\theta(x-x')p_1\tilde{q}_3\alpha_1 + \theta(x'-x)p_4\tilde{q}_2\beta_4\right] \\
& + uv\mathrm{e}^{i(x-x')k^{\mathrm{S}}_h}\left[\theta(x-x')r_1\tilde{s}_3\alpha_1 + \theta(x'-x)r_4\tilde{s}_2\beta_4\right] \\
& + v^2\mathrm{e}^{i(k^{\mathrm{S}}_h x - k^{\mathrm{S}}_e x')}\left[\theta(x-x')r_1\tilde{q}_3\alpha_1 + \theta(x'-x)r_4\tilde{q}_2\beta_4\right], \\
\left[G^r_{\mathrm{S_L}}(x,x')\right]_{eh,\downarrow\uparrow} = {}& u^2 \mathrm{e}^{-i(k^{\mathrm{S}}_e x - k^{\mathrm{S}}_h x')}\left[\theta(x-x')q_2\tilde{r}_4\alpha_4 + \theta(x'-x)q_3\tilde{r}_1\beta_1\right] \\
& + uv\mathrm{e}^{-i(x-x')k^{\mathrm{S}}_e}\left[\theta(x-x')q_2\tilde{p}_4\alpha_4 + \theta(x'-x)q_3\tilde{p}_1\beta_1\right] \\
& + uv\mathrm{e}^{-i(x-x')k^{\mathrm{S}}_h}\left[\theta(x-x')s_2\tilde{r}_4\alpha_4 + \theta(x'-x)s_3\tilde{r}_1\beta_1\right] \\
& + v^2\mathrm{e}^{-i(k^{\mathrm{S}}_h x - k^{\mathrm{S}}_e x')}\left[\theta(x-x')s_2\tilde{p}_4\alpha_4 + \theta(x'-x)s_3\tilde{p}_1\beta_1\right],
\end{aligned}
\tag{S20}
$$

where all the coefficients are known. In particular, as discussed in previous Section, $p_{1,4}, \tilde{p}_{1,4}, s_{2,3}, \tilde{s}_{2,3}, r_{1,4}, \tilde{r}_{1,4}$ are obtained after matching the scattering states $\Psi_i$ from Eqs. (S1), and their conjugated counterparts $\tilde{\Psi}_i$, at the interfaces using Eq. (S4). Moreover, as also explained at the beginning of this section, the coefficients $\alpha_i$ and $\beta_i$ are obtained by using Eq. (S13). In what follows we again focus on the local pair amplitudes and setting $x = x'$ we arrive at

$$
\begin{aligned}
\left[G^r_{\mathrm{S_L}}(x)\right]_{eh,\uparrow\downarrow} &= A_1 + A_2\mathrm{e}^{i(k^{\mathrm{S}}_e - k^{\mathrm{S}}_h)x} + A_3\mathrm{e}^{i(k^{\mathrm{S}}_h - k^{\mathrm{S}}_e)x}, \\
\left[G^r_{\mathrm{S_L}}(x)\right]_{eh,\downarrow\uparrow} &= B_1 + B_2\mathrm{e}^{-i(k^{\mathrm{S}}_e - k^{\mathrm{S}}_h)x} + B_3\mathrm{e}^{-i(k^{\mathrm{S}}_h - k^{\mathrm{S}}_e)x},
\end{aligned}
\tag{S21}
$$

where

$$
\begin{aligned}
A_1 &= uv\left[p_1\tilde{q}_3 + r_1\tilde{s}_3\right]\alpha_1, \\
A_2 &= u^2 p_1\tilde{s}_3\alpha_1, \\
A_3 &= v^2 r_1\tilde{q}_3\alpha_1, \\
B_1 &= uv\left[q_2\tilde{p}_4 + s_2\tilde{r}_4\right]\alpha_4, \\
B_2 &= u^2 q_2\tilde{r}_4\alpha_4, \\
B_3 &= v^2 s_2\tilde{p}_4\alpha_4,
\end{aligned}
\tag{S22}
$$

with,

$$
\begin{aligned}
\left[p_1\tilde{q}_3 + r_1\tilde{s}_3\right]\alpha_1 &= \frac{i\left[v^2\mathrm{e}^{2ik^{\mathrm{S}}_e L_{\mathrm{S}}}\left(u^2 - v^2\mathrm{e}^{i\phi}\right) + u^2\mathrm{e}^{2ik^{\mathrm{S}}_h L_{\mathrm{S}}}\left(-v^2 + u^2\mathrm{e}^{i\phi}\right)\right]}{v_{\mathrm{F}}(u^2-v^2)\left[-2u^2v^2\left(-1+e^{i\phi}\right)e^{iL_{\mathrm{S}}(k^{\mathrm{S}}_e + k^{\mathrm{S}}_h)} - u^2v^2 e^{2ik^{\mathrm{S}}_e L_{\mathrm{S}}} + v^4 e^{i(2k^{\mathrm{S}}_e L_{\mathrm{S}} + \phi)} - u^2v^2 e^{2ik^{\mathrm{S}}_h L_{\mathrm{S}}} + u^4 e^{i(2k^{\mathrm{S}}_h L_{\mathrm{S}} + \phi)}\right]}, \\
\left[q_2\tilde{p}_4 + s_2\tilde{r}_4\right]\alpha_4 &= \frac{i\left[u^2v^2 e^{i\phi}\left(e^{2ik^{\mathrm{S}}_h L_{\mathrm{S}}} - e^{2ik^{\mathrm{S}}_e L_{\mathrm{S}}}\right) + v^4 e^{2ik^{\mathrm{S}}_e L_{\mathrm{S}}} + u^4\left(-e^{2ik^{\mathrm{S}}_h L_{\mathrm{S}}}\right)\right]}{v_{\mathrm{F}}(u^2-v^2)\left[-u^2v^2 e^{i\phi}\left(e^{ik^{\mathrm{S}}_e L_{\mathrm{S}}} - e^{ik^{\mathrm{S}}_h L_{\mathrm{S}}}\right)^2 - 2u^2v^2 e^{iL_{\mathrm{S}}(k^{\mathrm{S}}_e + k^{\mathrm{S}}_h)} + v^4 e^{2ik^{\mathrm{S}}_e L_{\mathrm{S}}} + u^4 e^{2ik^{\mathrm{S}}_h L_{\mathrm{S}}}\right]}, \\
p_1\tilde{s}_3\alpha_1 &= \frac{iuv\left[e^{2ik^{\mathrm{S}}_h L_{\mathrm{S}}}\left(-v^2 + u^2 e^{i\phi}\right) - v^2\left(-1 + e^{i\phi}\right)e^{iL_{\mathrm{S}}(k^{\mathrm{S}}_e + k^{\mathrm{S}}_h)}\right]}{v_{\mathrm{F}}(u^2-v^2)\left[-2u^2v^2\left(-1+e^{i\phi}\right)e^{iL_{\mathrm{S}}(k^{\mathrm{S}}_e + k^{\mathrm{S}}_h)} - u^2v^2 e^{2ik^{\mathrm{S}}_e L_{\mathrm{S}}} + v^4 e^{i(2k^{\mathrm{S}}_e L_{\mathrm{S}} + \phi)} - u^2v^2 e^{2ik^{\mathrm{S}}_h L_{\mathrm{S}}} + u^4 e^{i(2k^{\mathrm{S}}_h L_{\mathrm{S}} + \phi)}\right]}, \\
r_1\tilde{q}_3\alpha_1 &= \frac{iuv e^{ik^{\mathrm{S}}_e L_{\mathrm{S}}}\left[u^2 e^{ik^{\mathrm{S}}_e L_{\mathrm{S}}} - v^2 e^{i(k^{\mathrm{S}}_e L_{\mathrm{S}} + \phi)} + u^2 e^{i(k^{\mathrm{S}}_h L_{\mathrm{S}} + \phi)} - u^2 e^{ik^{\mathrm{S}}_h L_{\mathrm{S}}}\right]}{v_{\mathrm{F}}(u^2-v^2)\left[-2u^2v^2\left(-1+e^{i\phi}\right)e^{iL_{\mathrm{S}}(k^{\mathrm{S}}_e + k^{\mathrm{S}}_h)} - u^2v^2 e^{2ik^{\mathrm{S}}_e L_{\mathrm{S}}} + v^4 e^{i(2k^{\mathrm{S}}_e L_{\mathrm{S}} + \phi)} - u^2v^2 e^{2ik^{\mathrm{S}}_h L_{\mathrm{S}}} + u^4 e^{i(2k^{\mathrm{S}}_h L_{\mathrm{S}} + \phi)}\right]}, \\
q_2\tilde{r}_4\alpha_4 &= \frac{iuv\left[-u^2 e^{i(L_{\mathrm{S}}(k^{\mathrm{S}}_e + k^{\mathrm{S}}_h) + \phi)} + u^2 e^{iL_{\mathrm{S}}(k^{\mathrm{S}}_e + k^{\mathrm{S}}_h)} + u^2 e^{i(2k^{\mathrm{S}}_e L_{\mathrm{S}} + \phi)} - v^2 e^{2ik^{\mathrm{S}}_e L_{\mathrm{S}}}\right]}{v_{\mathrm{F}}(u^2-v^2)\left[-u^2v^2 e^{i\phi}\left(e^{ik^{\mathrm{S}}_e L_{\mathrm{S}}} - e^{ik^{\mathrm{S}}_h L_{\mathrm{S}}}\right)^2 - 2u^2v^2 e^{iL_{\mathrm{S}}(k^{\mathrm{S}}_e + k^{\mathrm{S}}_h)} + v^4 e^{2ik^{\mathrm{S}}_e L_{\mathrm{S}}} + u^4 e^{2ik^{\mathrm{S}}_h L_{\mathrm{S}}}\right]}, \\
s_2\tilde{p}_4\alpha_4 &= \frac{iuv e^{ik^{\mathrm{S}}_h L_{\mathrm{S}}}\left[v^2 e^{i(k^{\mathrm{S}}_e L_{\mathrm{S}} + \phi)} - v^2 e^{ik^{\mathrm{S}}_e L_{\mathrm{S}}} + u^2 e^{ik^{\mathrm{S}}_h L_{\mathrm{S}}} - v^2 e^{i(k^{\mathrm{S}}_h L_{\mathrm{S}} + \phi)}\right]}{v_{\mathrm{F}}(u^2-v^2)\left[-u^2v^2 e^{i\phi}\left(e^{ik^{\mathrm{S}} L_{\mathrm{S}}} - e^{ik^{\mathrm{S}}_h L_{\mathrm{S}}}\right)^2 - 2u^2v^2 e^{iL_{\mathrm{S}}(k^{\mathrm{S}}_e + k^{\mathrm{S}}_h)} + v^4 e^{2ik^{\mathrm{S}}_e L_{\mathrm{S}}} + u^4 e^{2ik^{\mathrm{S}}_h L_{\mathrm{S}}}\right]}.
\end{aligned}
\tag{S23}
$$

Given the expressions above, we realize that the pair amplitudes in S acquire a lot more complicated form than in the outer N regions, but it is still possible to isolate their symmetries. Taking into account that Eqs. (S21) correspond to local pair amplitudes, which have even-parity symmetry, we decompose their spin symmetry following Eq. (S15) and

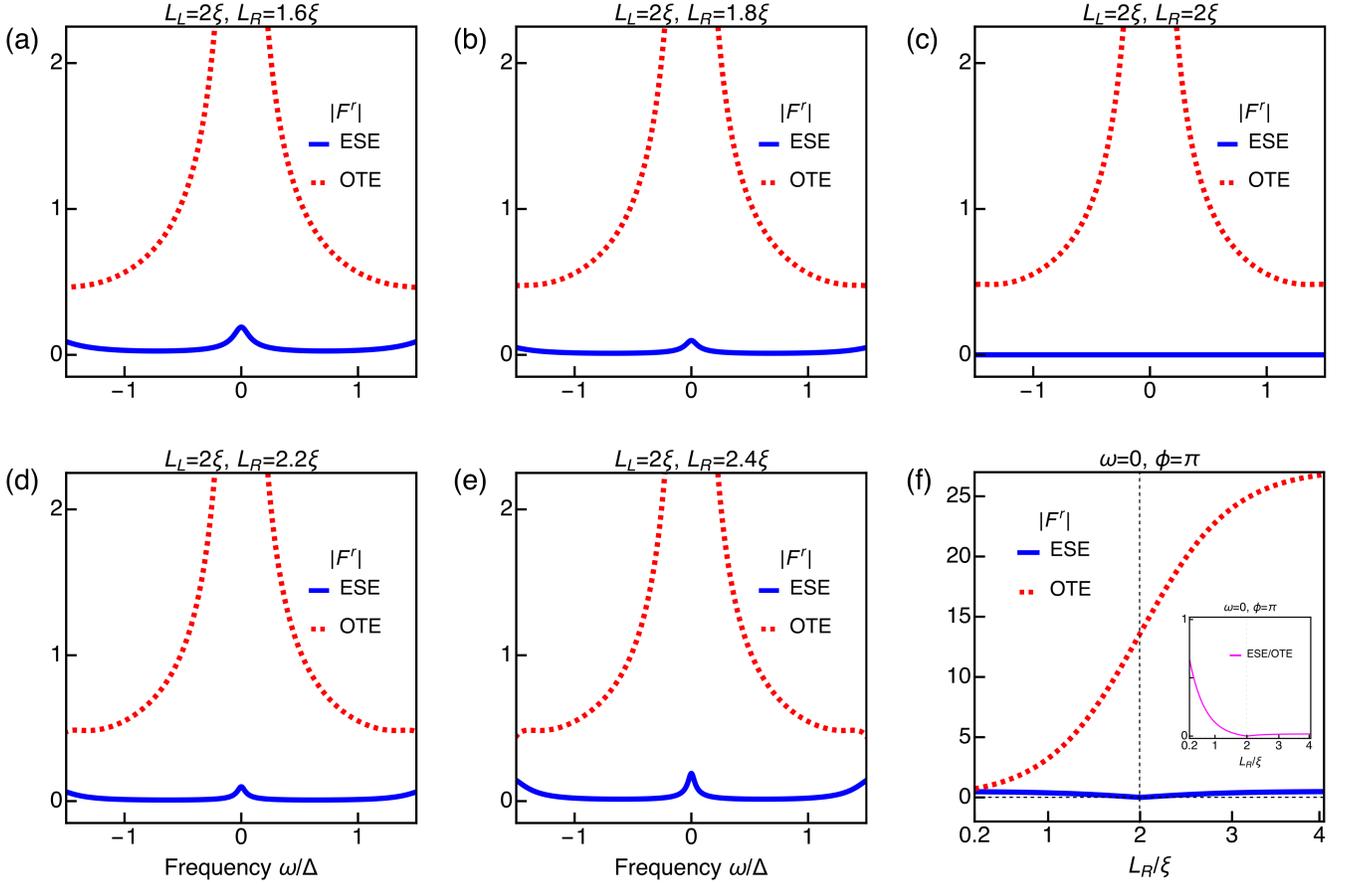

FIG. S2. (a-e) Absolute values of the ESE (blue) and OTE (red) pair amplitudes as a function of frequency at $\phi = \pi$, $L_\mathrm{L} = 2\xi$ for distinct values of $L_\mathrm{R}$. (f) Absolute values of the ESE (blue), OTE (red) pair amplitudes, and their ratio (inset) as a function of $L_\mathrm{R}$ at $\omega = 0$, $\phi = \pi$. Parameters: $x = L_\mathrm{L}$, $\mu = v_\mathrm{F} = \Delta = 1$, $L_\mathrm{L} = 2\xi$.

obtain

$$F_{0(3)}^r(x,\omega) = \frac{A_1 \pm B_1}{2} + \frac{A_2 \pm B_3}{2} \mathrm{e}^{i(k_e^\mathrm{S} - k_h^\mathrm{S})x} + \frac{A_3 \pm B_2}{2} \mathrm{e}^{i(k_h^\mathrm{S} - k_e^\mathrm{S})x}, \qquad (\text{S24})$$

where $A_{1,2,3}$ and $B_{1,2,3}$ are given by Eqs. (S22). The two pair amplitudes, $F_{0,3}^r$, correspond to the ESE (OTE) symmetries, which generally coexist in the junction. We note that the first term in Eq. (S24) is space independent and thus exists throughout the S regions. The second and third terms, however, depend on the spatial coordinate $x$ and thus exhibit a strong decay from the interfaces for frequencies below the gap $|\omega| < \Delta$. These two latter terms can be viewed as interface-induced components of the pair amplitudes. The obtained ESE and OTE pair amplitudes in Eqs. (S24) are discussed in the section "Induced odd-$\omega$ pairs" of the main text and their interesting behavior is presented in Fig. 3 of the main text.

### Impact of S region asymmetry on the pair amplitudes

Here we discuss the impact of having distinct S regions on the induced local ESE and OTE pair amplitudes. For this purpose, we follow the same steps as discussed in the previous subsection and obtain pair amplitudes that acquire the same expressions as Eqs. (S21) but with the difference that now the coefficients $A_i$ and $B_i$ depend on the two different lengths of the S regions. As the expressions for the coefficients become more complicated, it is here simpler to plot the ESE and OTE components for visualization. In Fig. S2(a-e) we plot the absolute values of the ESE (blue) and OTE (red) pair amplitudes as a function of frequency at $x = L_\mathrm{L}$ for $\phi = \pi$, $L_\mathrm{L} = 2\xi$ at multiple different values of $L_\mathrm{R}$. The main observation is that, even though some asymmetry in the lengths of S favors the emergence of a small finite ESE pairing component at $x = L_\mathrm{L}$, $\omega = 0$, $\phi = \pi$, the OTE component is still by far the largest amplitude. This

is a consequence of the zero-frequency protected crossing $\phi = \pi$ which signal the formation of a pair of topological ABSs. The OTE component always grows by increasing $L_R$, seen Fig. S2(f). In contrast, the ESE first decreases, then vanishes at $L_L = L_R$, followed by a slight increase. All the ESE amplitudes, however, are much smaller than the OTE amplitudes, as explicitly seen when extracting their ratio in the inset in Fig. S2(f)]. This shows that the dominating OTE and near vanishing ESE pair amplitudes are present also in non-ideal and slightly asymmetric junctions.

---


	
\end{document}